\documentclass{appolb}
\usepackage{epsfig}
\usepackage{graphics, color, cite, curves, amssymb, amsmath, rotating, epic}
\usepackage{eepic}
\newcommand{\gaz}{g_A^{\mbox{$\scriptscriptstyle (Z)$}}}
\newcommand{\run}[1]{\widetilde{\alpha}_{#1}}
\newcommand{\hsp}[1]{\hspace*{#1 mm}}
\newcommand{\smallfrac}[2]{\mbox{\small ${\displaystyle \frac{#1}{#2}}$}}
\newcommand{\footfrac}[2]%

\begin{document}

\title{
Probing polarized strangeness in the proton: \\
the use of heavy quarks and the renormalization group 
%%and heavy quarks %
\thanks{Presented at Cracow Epiphany Conference 2003: Heavy-flavours}%
}
\author{Steven D. Bass
\address{
High Energy Physics Group, \\
Institute for Experimental Physics 
and Institute for Theoretical Physics, 
\\
Universit\"at Innsbruck, \\
Technikerstrasse 25, A 6020 Innsbruck, Austria}
%\and
%the Name(s) of other Author(s)
%\address{and their affiliation}
}
\maketitle
\begin{abstract}
Once heavy-quark corrections are take into account $\nu p$ elastic
scattering provides a complementary probe of polarized strangeness
in the nucleon to measurements from inclusive and semi-inclusive
polarized deep inelastic scattering.
We review the different types of experiment and a 
recent NLO calculation
of heavy-quark contributions to the weak neutral-current 
axial-charge (measureable in $\nu p$ elastic scattering) 
performed using Witten's heavy-quark renormalization group method.
\end{abstract}
\PACS{
11.10.Hi, 12.15.Mm, 12.38.Cy, 12.39.Hg, 13.60.Hb,13.88.+e}

\section{Introduction}

Understanding the internal spin structure of the proton is one of the most 
challenging problems facing subatomic physics: 
How is the spin of the proton built up out from the intrinsic spin 
and orbital angular momentum of its quark and gluonic constituents ? 
A key issue is the contribution of polarized strangeness
in building up the spin of the proton.
Fully inclusive and semi-inclusive measurements of polarized deep 
inelastic scattering together with elastic $\nu p$ scattering
provide complemetary information about the transverse momentum and
Bjorken $x$ distributions of strange quark polarization in the proton.
In this paper we review the different experiments and 
outline the vital role of charm (and other heavy quarks) 
in helping to pin down the size and interpretation of polarized 
strangeness in the proton.

Our present knowledge about the spin structure of the nucleon comes from 
polarized deep inelastic scattering. 
Following pioneering experiments at SLAC \cite{slac}, recent experiments 
in fully inclusive polarized deep inelastic scattering have extended 
measurements of the nucleon's $g_1$ spin dependent structure function 
to lower values of Bjorken $x$ where the nucleon's sea becomes important 
\cite{windm}.
From the first moment of $g_1$, these experiments have revealed a small 
value for the flavour-singlet axial-charge:
\begin{equation}
g_A^{(0)}\bigr|_\mathrm{inv} = 0.2 - 0.35 
\label{inv}\end{equation}
%\begin{equation}
%g_A^{(0)}\bigr|_\mathrm{pDIS} = \Delta u + \Delta d + \Delta s = 0.2 - 0.35 .
%\label{inv}
%\end{equation}
This result is particularly interesting \cite{spinrev} because $g_A^{(0)}$ is 
interpreted in the parton model as the fraction of the proton's spin which is 
carried 
by the intrinsic spin of its quark and antiquark constituents.
The value (\ref{inv}) is about half the prediction of 
relativistic constituent quark models ($\sim 60\%$).
It corresponds to a negative strange-quark polarization
\begin{equation}
\Delta s = -0.10 \pm 0.04 .
\label{deltas}
\end{equation}
(polarized in the opposite direction to the spin of the proton).

The small value of $g_A^{(0)}$ measured in polarized deep inelastic scattering 
has inspired vast experimental and theoretical activity to understand the spin 
structure of the proton. New experiments are underway or being planned 
to map out the proton's spin-flavor structure and to measure the amount of 
spin carried by polarized gluons in the polarized proton. 
These include semi-inclusive polarized deep inelastic scattering \cite{miller},
polarized proton-proton collisions at the Relativistic Heavy Ion Collider 
(RHIC) \cite{rhic}, and polarized $ep$ collider studies \cite{bassdr}.
It is essential to ensure that the theory and experimental acceptance are 
correctly matched when extracting new information from present and future
exeriments \cite{bassa}.
For example, the spin-flavour structure of the sea extracted from 
semi-inclusive measurements of polarized deep inelastic 
scattering may depend strongly on the angular acceptance of the detector.

A clean measurement of parity violating $\nu p$ elastic scattering 
\cite{garvey,tayloe} would
provide an exciting new opportunity to probe strangeness 
polarization in the proton.
This experiment measures the weak axial charge $\gaz$ 
through elastic Z$^0$ exchange.
Because of anomaly cancellation in the Standard Model 
the weak neutral current couples to the combination $u-d+c-s+t-b$, 
{\it viz.}
\begin{equation}
J_{\mu5}^Z\ 
=\ \smallfrac{1}{2} \biggl\{\,\sum_{q=u,c,t} - \sum_{q=d,s,b}\,\biggr\}\:
        \bar{q}\gamma_\mu\gamma_5q
\label{g}\end{equation}
It measures the combination:
\begin{equation}
2\gaz = \bigl( \Delta u - \Delta d - \Delta s \bigr) 
       + \bigl( \Delta c - \Delta b + \Delta t \bigr)
\label{g1}
\end{equation}
where $\Delta q$ refers to the expectation value
\[ \langle p,s|\,\bar{q}\gamma_\mu\gamma_5q\,|p,s \rangle 
  = 2m_p s_\mu\Delta q \]
for a proton of spin $s_\mu$ and mass $m_p$. 
The contribution $\Delta u - \Delta d$ in Eq.(\ref{g1}) 
is just the axial charge measured 
in neutron beta-decays ($g_A^{\scriptscriptstyle (3)} = 1.267 \pm 0.004$). 
Hence,
once heavy-quark corrections \cite{CWZ,KM,Bass,bcstnup} 
have been taken into account, 
$\gaz$ is related (modulo the issue of $\delta$--function 
terms at $x=0$ \cite{topology}) 
to the strange-quark axial-charge
(polarized strangeness), 
defined scale invariantly,
which is extracted from polarized deep inelastic scattering.
A quality $\nu p$ elastic measurement
would be independent of assumptions about 
the $x \sim 0$ behaviour of the proton's $g_1$ spin structure function.
A definitive measurement of $\nu p$ elastic scattering 
may be possible using the miniBooNE set-up at FNAL \cite{tayloe}.

We next review the different types of experiment: inclusive and
semi-inclusive polarized deep inelastic scattering, and elastic $\nu p$ 
scattering respectively (Sections 2-4).
The renormalization scale invariant axial-charge $\Delta q|_{\rm inv}$
is defined in Section 2.
Section 4 summarizes the recent NLO calculation \cite{bcstnup} of the 
heavy-quark contributions to $g_A^{(Z)}$, which was performed using the 
rigor of Witten's heavy-quark renormalization group \cite{witten,bcstb}.

\section{Polarized deep inelastic scattering}

The value of $g_A^{(0)}|_{\rm inv}$ extracted from polarized 
deep inelastic
scattering is obtained as follows. 
The first moment of  the structure function $g_1$ 
to the scale-invariant axial charges of the target nucleon by:
\begin{eqnarray}
\int_0^1 dx \ g_1^p (x,Q^2) &=&
\Biggl( {1 \over 12} g_A^{(3)} + {1 \over 36} g_A^{(8)} \Biggr)
\Bigl\{1 + \sum_{\ell\geq 1} c_{{\rm NS} \ell\,}
\alpha_s^{\ell}(Q)\Bigr\} \nonumber \\
&+& {1 \over 9} g_A^{(0)}|_{\rm inv}
\Bigl\{1 + \sum_{\ell\geq 1} c_{{\rm S} \ell\,}
\alpha_s^{\ell}(Q)\Bigr\}
\ + \ {\cal O}({1 \over Q^2}).
\end{eqnarray}
Here $g_A^{(3)}$, $g_A^{(8)}$ and $g_A^{(0)}|_{\rm inv}$ are 
the isotriplet, SU(3)
octet and scale-invariant  flavour-singlet axial charges respectively. 
The flavour non-singlet $c_{{\rm NS} \ell}$ and singlet 
$c_{{\rm S} \ell}$ 
Wilson coefficients are calculable in $\ell$-loop perturbative QCD.
The main source of experimental error on deep inelastic measurements
of $g_A^{(0)}|_{\rm inv}$ comes from the extrapolation of the measured
$g_1$ spin structure function to lower values of Bjorken $x$.

Note that the first moment of $g_1$ is constrained by low energy weak 
interactions.
For proton states $|p,s\rangle$ with momentum $p_\mu$ and spin $s_\mu$
\begin{eqnarray}
2 m s_{\mu} \ g_A^{(3)} &=&
\langle p,s | 
\left(\bar{u}\gamma_\mu\gamma_5u - \bar{d}\gamma_\mu\gamma_5d \right)
| p,s \rangle   \nonumber \\
2 m s_{\mu} \ g_A^{(8)} &=&
\langle p,s |
\left(\bar{u}\gamma_\mu\gamma_5u + \bar{d}\gamma_\mu\gamma_5d
                   - 2 \bar{s}\gamma_\mu\gamma_5s\right)
| p,s \rangle 
\end{eqnarray}
Here $g_A^{\scriptscriptstyle (3)} = 1.267 \pm 0.004$ 
is the isotriplet axial charge measured in neutron beta-decay;
$g_A^{\scriptscriptstyle (8)} = 0.58 \pm 0.03$ 
is the octet charge measured independently in hyperon beta decay
\cite{fecrgr}.

The scale-invariant flavour-singlet axial charge $g_A^{(0)}|_{\rm inv}$ 
is defined as follows.
Let $\alpha_f = g_f^2/4\pi$ and $\beta_f(\alpha_f)$ be the gluon coupling
and beta function for $\overline{\mbox{\small MS}}$ renormalized 
quantum chromodynamics (QCD) with $f$ flavours and $N_c=3$ colours, 
and let  $\gamma_f(\alpha_f)$ 
be the gamma function for the singlet current
\begin{equation} 
\bigl(\bar{u}\gamma_\mu\gamma_5u + \bar{d}\gamma_\mu\gamma_5d 
      + \ldots\bigr)_f 
 = \sum_{k=1}^f\bigl(\bar{q}_k\gamma_\mu\gamma_5q_k\bigr)_f 
\label{singlet}\end{equation}
which is induced by the QCD axial anomaly
\begin{equation}
\partial^{\mu} 
\sum_{k=1}^f\bigl(\bar{q}_k\gamma_\mu\gamma_5q_k\bigr)_f
= 
2 f \partial^{\mu} K_{\mu} + \sum_{i=1}^f 2i m_i {\bar q}_i \gamma_5 q_i
\end{equation}
where $K_{\mu}$ is the gluonic Chern-Simons current.
A scale-invariant current $(S_{\mu 5})_f$ is obtained 
when 
(\ref{singlet}) is multiplied by 
\begin{equation}
E_f(\alpha_f)\, =\, \exp\!\int^{\alpha_f}_0\hsp{-2.5} dx\,
                      \frac{\gamma_f(x)}{\beta_f(x)}
\label{f0}\end{equation}
%%Up to $O(m_h^{-1})$ corrections, 
The invariant singlet charge is given by
\begin{align}
g_A^{(0)}\bigr|_\mathrm{inv} &=  E_3(\alpha_3) 
       \bigl(\Delta u + \Delta d + \Delta s\bigr)_3   \nonumber \\
&= \bigl(\Delta u + \Delta d + \Delta s\bigr)_\mathrm{inv}
\label{inv2}\end{align}
Flavour-dependent, scale-invariant axial charges\, $\Delta q|_{\rm inv}$ 
such as 
\begin{equation}
\Delta s|_\mathrm{inv} 
 = \smallfrac{1}{3}\Bigl(g_A^{(0)}\bigr|_\mathrm{inv}-g_A^{(8)}\Bigr)
\end{equation}
can then  be obtained from linear combinations of (\ref{inv2}) and
\begin{gather}
g_A^{(3)} = \Delta u - \Delta d  
          = \bigl(\Delta u - \Delta d\bigr)_\mathrm{inv}   \nonumber \\
g_A^{(8)} = \Delta u + \Delta d - 2 \Delta s  
          = \bigl(\Delta u + \Delta d - 2 \Delta s\bigr)_\mathrm{inv}
\end{gather}
Modulo heavy-quark corrections, 
$g_A^{(3)}$ and $g_A^{(8)}$ 
together with 
$g_A^{(Z)}$ would
provide a weak interaction determination of $\Delta s|_{\rm inv}$,
complementary to the DIS measurement.

\section{Semi-inclusive polarized deep inelastic scattering}

Semi-inclusive measurements of fast pions and kaons in the current 
fragmentation region with final state particle identification can
be used to reconstruct the individual up, down and strange quark 
contributions to the proton's spin \cite{close,closem}.
In contrast to inclusive polarized deep inelastic scattering 
where the $g_1$ structure function is deduced by detecting only 
the scattered lepton, the detected particles in the semi-inclusive 
experiments are high-energy (greater than 20\% of the energy of the 
incident photon)
charged pions and kaons in coincidence with the scattered lepton.
For large energy fraction $z=E_h/E_{\gamma} \rightarrow 1$
the most probable occurence is that the detected $\pi^{\pm}$ and 
$K^{\pm}$
contain the struck quark or antiquark in their valence Fock state. 
They therefore act as a tag of the flavour of the struck quark.

New semi-inclusive data reported 
by the HERMES experiment \cite{hermessemi,miller} (following earlier 
work by SMC \cite{smcsemi}) 
suggest that the light-flavoured (up and down) sea measured in these 
semi-inclusive experiments contributes close to zero to the proton's 
spin.
Furthermore, recent HERMES data \cite{miller} also tends to 
favour slightly {\it positively} polarized strangeness in the 
kinematical range probed by the experiment.

An important issue for semi-inclusive measurements is 
the angular coverage of the detector \cite{bassa}. 
The non-valence spin-flavor structure of the proton extracted from 
semi-inclusive measurements of polarized deep inelastic scattering 
may depend strongly on the transverse momentum (and angular) 
acceptance of the detected final-state hadrons which are used 
to determine the individual polarized sea distributions. 
The present semi-inclusive experiments detect final-state hadrons
produced only at small angles from the incident lepton beam (about 
150 mrad angular coverage) whereas 
the perturbative QCD 
``polarized gluon interpretation'' \cite{etar} of 
the inclusive measurement (\ref{deltas}) involves physics at 
the maximum transverse momentum \cite{ccm,bint} and large angles.

New semi-inclusive measurements with increased luminosity and a 
$4 \pi$ detector, as proposed 
for the next generation Electron Ion Collider facility 
in the United States, would be extremely useful to map out 
the transverse momentum distribution of the total polarized 
strangeness (\ref{deltas}) measured in inclusive deep inelastic scattering.

\begin{figure}[h] 
\includegraphics{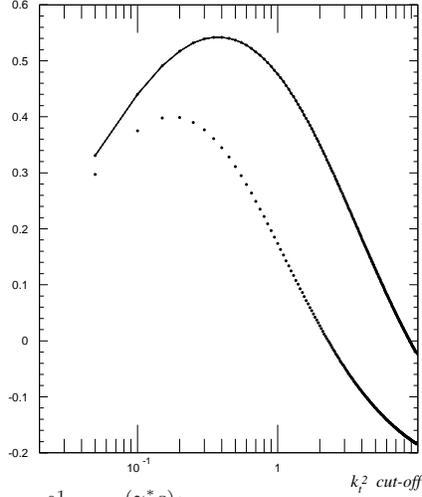} %\label{f1a}
\begin{center} 
\vspace{6.5cm} 
\parbox{12.0cm} 
{\caption[Delta]
{
$\int_0^1 dx \ g_1^{(\gamma^* g)}|_{\rm soft}$ 
for polarized strangeness production 
with $k_t^2 < \lambda^2$
in units of ${\alpha_s \over 2 \pi}$.
Here 
$Q^2=2.5$GeV$^2$ (dotted line) and 10GeV$^2$ (solid line).
}
\label{fig1}} 
\end{center} 
\end{figure}
\begin{figure}[h] 
\includegraphics{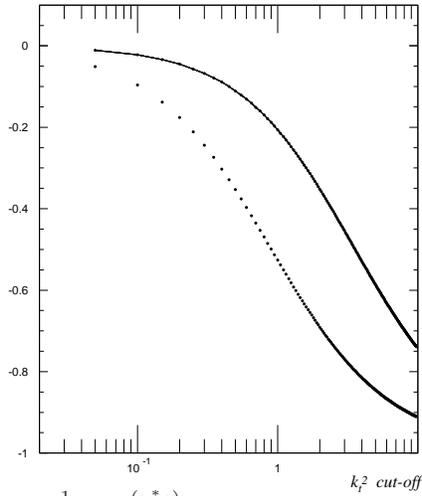} %\label{f1a}
\begin{center} 
\vspace{6.5cm} 
\parbox{12.0cm} 
{\caption[Delta]
{
$\int_0^1 dx \ g_1^{(\gamma^* g)}|_{\rm soft}$ 
for light-flavor ($u$ or $d$) production 
with $k_t^2 < \lambda^2$ 
in units of ${\alpha_s \over 2 \pi}$.
Here 
$Q^2=2.5$GeV$^2$ (dotted line) and 10GeV$^2$ (solid line).
}
\label{fig2}} 
\end{center} 
\end{figure}

To understand this physics, consider the polarized photon-gluon fusion 
contribution to the polarized sea \cite{ccm,bint}.  
In leading twist, the first moment of the $g_1$ spin structure function 
for polarized photon-gluon fusion $(\gamma^* g \rightarrow q {\bar q})$ 
receives a positive contribution proportional to the mass squared of the 
struck quark or antiquark which originates from low values of quark 
transverse momentum, $k_t$, with respect to the photon-gluon direction.
It also receives a negative contribution from $k_t^2 \sim Q^2$, where 
$Q^2$ is the virtuality of the hard photon.
Thus, the spin-flavor structure of the sea extracted from semi-inclusive 
measurements depends strongly on the $k_t$ distribution of the detected 
hadrons.
The positive mass-dependent contribution from low $k_t^2$ can safely be 
neglected for light-quark flavour (up and down) production.  
It is very important for strangeness 
(and charm \cite{bbs2,fms}) production. 
Let $P^2$ and denote virtuality of 
the target gluon and $m$ denote the mass of the struck quark.
The fully inclusive first moment 
$\int_0^1 dx g_1^{(\gamma^* g)}$ 
is equal to $-{\alpha_s \over 2 \pi}$ 
in the limit $m^2 \ll P^2$ and vanishes in the limit $m^2 \gg P^2$.
The vanishing of
 $\int_0^1 dx g_1^{(\gamma^* g)}$ in the limit $m^2 \ll P^2$
 to leading order in $\alpha_s (Q^2)$
 follows from an application \cite{bassbs} of 
 the fundamental Drell-Hearn-Gerasimov sum-rule \cite{drellh}.

The practical consequence \cite{bassa}
of the strange quark mass on polarized
photon-gluon fusion is shown in Figs. 1-2. 
Here we let 
$g_1^{(\gamma^* g)}|_{\rm soft} (\lambda)$ denote the contribution
to 
$g_1^{(\gamma^* g)}$ for photon-gluon fusion where the hard photon 
scatters on the struck quark or antiquark carrying 
transverse momentum $k_t^2 < \lambda^2$.
Figs. 1 and 2 show the first moment of $g_1^{(\gamma^* g)}|_{\rm soft}$ 
for the strange and light (up and down) flavour production 
respectively as a function of the transverse momentum cut-off $\lambda^2$. 
Here we set $Q^2 =2.5$GeV$^2$ 
(corresponding to the HERMES experiment) and 10GeV$^2$ (SMC). 
Following \cite{ccm}, 
we take $P^2 \sim \Lambda_{\rm qcd}^2$ and set $P^2 = 0.1$GeV$^2$.
Observe the small value for the light-quark 
sea polarization at low transverse momentum and 
the positive value for the integrated 
strange sea polarization at low $k_t^2$:
$k_t < 1.5$GeV at the HERMES $Q^2=2.5$GeV$^2$.
When we relax the cut-off, increasing the acceptance of the experiment,
the measured strange sea polarization changes sign and becomes 
negative (the result implied by fully inclusive deep inelastic measurements).
Note that for $\gamma^*g$ fusion the cut-off $k_t^2 < \lambda^2$ 
is equivalent to a cut-off on the angular acceptance 
$\sin^2 \theta < 4 \lambda^2 / \{s - 4m^2 \}$ 
where $\theta$ is defined relative to the photon-gluon direction
and $s$ is the centre of mass energy for the photon-gluon collision.
Leading-twist negative sea polarization at 
$k_t^2 \sim Q^2$ corresponds, in part, 
to final state hadrons produced at large angles.
For HERMES the average transverse momentum of the detected 
final-state fast 
hadrons is less than about 0.5 GeV whereas for SMC 
the $k_t$ of the detected fast pions was less than about 1 GeV.

Semi-inclusive measurements of charm production in polarized 
lepton-proton scattering will be made by the 
COMPASS experiment at CERN and the SLAC experiment E-161.
The aim of these experiments is to shed new light on the 
role of polarized glue in the nucleon, which will, in turn, 
help to resolve the origin of the negative polarized strange 
sea (\ref{deltas}) 
extracted from inclusive polarized deep inelastic scattering.

\section{$\nu p$ elastic scattering}

The $\nu p$ elastic process \cite{garvey} measures the neutral current 
axial-charge $\gaz$, where 
\begin{equation}
2\gaz = \bigl( \Delta u - \Delta d - \Delta s \bigr) 
       + \bigl( \Delta c - \Delta b + \Delta t \bigr)
%%\label{g1}
\end{equation}
Bass, Crewther, Steffens and Thomas \cite{bcstnup} 
have recently combined
Witten's renormalization group \cite{witten,bcstb}
with the 
matching conditions of Bernreuther and Wetzel \cite{BW}
to calculate at 
next-to-leading order the complete heavy-quark contribution 
to $\gaz$.
One finds that, when first $t$, then $b$, and finally $c$ 
are decoupled from (\ref{g1}), the full NLO result is
\begin{equation}
2\gaz\, =\, \bigl(\Delta u - \Delta d - \Delta s\bigr)_\mathrm{inv}
           +\hsp{0.2} {\cal P}\hsp{0.1}\bigl(
               \Delta u + \Delta d + \Delta s\bigr)_\mathrm{inv} 
    +\, O(m_{t,b,c}^{-1})
\label{g2}\end{equation}
where ${\cal P}$ is a polynomial in the running couplings
$\run{h}$,
\begin{align}
{\cal P}\, =\, &\smallfrac{6}{23\pi}\bigl(\run{b}-\run{t}\bigr)
             \Bigl\{1 + \smallfrac{125663}{82800\pi}\run{b}
                      + \smallfrac{6167}{3312\pi}\run{t}
                      - \smallfrac{22}{75\pi}\run{c}  \Bigr\}
\nonumber \\
&\phantom{+\ \Biggl[} - \smallfrac{6}{27\pi} \run{c}
                      - \smallfrac{181}{648 \pi^2}\run{c}^2
                      + O\bigl(\run{t,b,c}^3\bigr)
\label{j3}
\end{align}
Here $(\Delta q)_\mathrm{inv}$ 
denotes the scale-invariant version of $\Delta q$ and $\run{h}$ 
denotes Witten renormalization group invariant running couplings. 
These Witten couplings \cite{witten} 
are defined for the full 
theory (including the heavy-quark $h$) as follows.
Let $m_h$ be the $\overline{\mbox{\small MS}}_F$ renormalized 
mass and $\alpha_F$ denote the $F$ flavour running coupling.
Then the Witten coupling
\begin{equation} 
\run{h} = \run{h}\bigl(\alpha_F, \ln(m_h/\bar{\mu})\bigr)
\end{equation} 
is defined via
\begin{equation} 
\ln(m_h/\bar{\mu})\ 
=\ \int^{\run{h}}_{\alpha_F}\! dx\,\bigl(1-\delta_F(x)\bigr)/\beta_F(x)
\label{f3} 
\end{equation}
where $\delta_F$ denotes the mass anomalous dimension and $\beta_F$ is
the $F$-flavour $\beta$ function.
It satisfies the constraints
\begin{equation} 
\run{h}(\alpha_F,0) = \alpha_F \hsp{3},\hsp{3}  
\run{h}(\alpha_F,\infty) = 0
\end{equation}
the latter being a consequence of 
the asymptotic freedom of the $F$ 
flavour theory ($F \leqslant 16$). 
The Witten coupling is renormalization group invariant:  
\vspace{-1mm}
\begin{equation} 
{\cal D}_F \run{h} = 0
\end{equation} 
Taking $\widetilde{\alpha}_t = 0.1$, 
$\widetilde{\alpha}_b = 0.2$ and $\widetilde{\alpha}_c = 0.35$ 
in (\ref{j3}), we find a small heavy-quark correction factor 
${\cal P}= -0.02$, with LO terms dominant.

These results are manifestly renormalization group invariant. 
They will permit a theoretically clean determination of the
strange-quark axial-charge, $\Delta s|_{\rm inv}$, from the
neutrino-proton elastic process. 
The results (\ref{g2},\ref{j3}) extend to NLO and 
make more precise the well 
known work of Collins, Wilczek and Zee \cite{CWZ} and 
Kaplan and Manohar \cite{KM},
where heavy-quark effective theory was used to estimate $\gaz$ 
in leading order (LO) for sequential decoupling of $t,b$ and $t,b,c$
respectively.

There is interest \cite{tayloe} to perform the experimental measurement 
at FNAL using the mini-BooNE set-up with very low duty factor neutrino 
beam to control background. 
The estimated error on the strange quark polarization one could extract 
from this experiment is 0.03, competitive with the error from the present 
polarized deep inelastic measurements.

\vspace{1.0cm}

\noindent{\bf Acknowledgements:}

SDB is supported by a Lise-Meitner Fellowship (M683) from the 
Austrian Science Fund (FWF).
It is a pleasure to thank
S. J. Brodsky, R. J. Crewther, I. Schmidt, F. M. Steffens and 
A. W. Thomas for collaboration on many of the results presented
here, and M. Jezabek for organizing this stimulating meeting.

\end{document}